\documentstyle[aps]{revtex}
\begin{document}
\draft
\date{\today} 
\title{TEN MAJOR CHALLENGES IN COSMOLOGY}
\author{Reuven Opher\thanks{email: opher@astro.iag.usp.br}}
\address{Departamento de Astronomia\\
Instituto de Astronomia, Geof\'\i sica e Ci\^encias Atmosf\'ericas\\
Universidade de S\~ao Paulo\\
Rua do Mat\~ao 1226, Cidade Universit\'aria\\
CEP 05508-900 S\~ao Paulo, S.P., Brazil}
\maketitle
\begin{abstract}
What are the important unanswered questions in cosmology? The choice as to which are the most important is somewhat personal. In the view of the author, the ten topics chosen here for discussion must be included among 
the most fundamental. Present cosmological theory is unable to fully describe the evolution of the universe from the time of the Big Bang and a lack of observations makes it difficult to verify theoretical predictions. The questions posed here will only be answered when we have both better theory and more observations.
\end{abstract}
\section{INTRODUCTION}
\label{Sec:intro}
\par
Cosmology has progressed considerably since 1916, when Einstein published his Theory of General Relativity. Then,
the observable universe was about a third of the size of our galaxy and we knew little or nothing about the history of the universe, compared to what we know today. Present cosmological theories of the evolution of the universe from the time of the Big Bang are supported by the following pillars.
\begin{enumerate}
\item{\bf Einstein's theory of general relativity}\\
Almost all cosmological theories are based on the Friedman-Robertson-Walker form of Einstein's 1916 Theory of General Relativity, in which homogeneity and isotropy of the universe are assumed. 
\par
\item{\bf Hubble's law of expansion}\\
In the late 1920's, Hubble discovered that the universe is expanding, implying a Big Bang origin about 10-20 billion years ago. The Hubble constant, $H_0\;(\equiv {\dot{a}}/a,$ where $a$ is the scale factor of the universe) 
is $\sim 70\;\rm{km\; s^{-1}\;{Mpc}^{-1}}.$ 
\par
\item{\bf The cosmic microwave background}\\
Penzias and Wilson discovered the cosmic microwave background (CMB), the remnant radiation of the Big Bang   
with a temperature $T\sim 2.7\;\rm{K},$ in the early 1960's.
\par
\item{\bf Primordial nucleosynthesis}\\
It was shown in the early 1960's that the $^4$He abundance in the universe was primarily primordial. In the 
1970's and 1980's, the abundances of D and $^3$He were also shown to be primarily primordial. This was extended 
to $^7$Li in the 1980's (see, e.g., Schramm and Turner \cite{SCT} for a review). The primordial nucleosynthesis 
of D, $^3$He, $^4$He, and $^7$Li indicates that $\Omega_{\rm{B}}$, the ratio of the baryon density of the 
universe to the critical density $\rho_{\rm{crit}},$ where $\rho_{\rm{crit}}=3H_0^2/8\pi G$ and $G$ is the
gravitational constant, is $\sim 4\%.$ 
\par
\item{\bf The present acceleration of the universe}\\
The present dimensionless acceleration of the universe is defined as $\ddot{a}/a H^2_0.$ It is found to be positive, as indicated by supernovae type Ia (SNIa) data \cite{PR,RI}.
\par
\item{\bf The density of  pressureless dark matter}\\
As indicated primarily by data from galaxy clusters (GC), $\Omega_{\rm{M}},$ 
the ratio of the density of pressureless dark matter to $\rho_{\rm{crit}},$ is $\sim 30\%.$  
\par
\item{\bf The total density of the universe}\\
From the CMB, SNIa, and GC data, $\Omega_{\rm{T}},$ the ratio of the total density of the universe to 
$\rho_{\rm{crit}},$ is $\sim 1.$ 
\par
\item{\bf The primordial density fluctuation spectrum}\\
Data from the CMB and the large scale structure of the universe (LSS) show that the primordial density 
fluctuation spectrum, $\delta\rho/\bar{\rho}$ vs size $\lambda,$ is approximately scale-invariant, where $\delta\rho\equiv \rho - \bar{\rho}$ and $\rho(\bar{\rho})$ is the density (average density) of the universe 
over a dimension $\lambda$. Scale invariance signifies that $\delta\rho/\bar{\rho}$ has a constant value 
$(\sim 10^{-5})$ when $\lambda=c\tau_U,$ where $\tau_U$ is the age of the universe when the fluctuations entered the horizon and $c$ is the velocity of light.
\end{enumerate}
\par
Although the above pillars have provided us with a great deal of knowledge on which to base our cosmological theories, we are still far from having a clear picture of just how the observed universe came about. The most popular cosmological theory is the Inflation Theory of Guth \cite{GU}, according to which, the universe passed through an exponential superluminal expansion epoch of $\geq 60$ $e$-foldings. This inflation scenario can 
explain the large number of CMB photons (large entropy) in the observable universe (third pillar), $\Omega_{\rm{T}} \sim 1$ (seventh pillar), and the approximately scale-invariant density fluctuation spectrum (eigth pillar). With this scenario, the assumption of the homogeneity and isotropy of the universe (included in the first pillar) can also be justified. However, it leaves various important questions still unanswered.
In order to explain the fourth pillar (baryon density), we require  a theory of baryogenesis, which doesn't yet exist. A complete theory of cosmology, which might incorporate the Inflation Theory as well as a theory of baryogenesis, must also explain the present acceleration of the universe (fifth pillar), $\Omega_M\sim 0.3$ 
(sixth pillar), as well as the formation and distribution of galaxies. The present Hubble expansion of the universe (second pillar) is generally assumed to be a consequence of the Big Bang, which is taken as a basic
premise. 
\par
This lecture is an invitation to reflect upon what is, in your judgement, the ten major challenges in cosmology today. Inevitably, many of the decisions that we, in the scientific community,  are called upon to make, are 
based on this judgement. The topics presented here are those that I consider to be the ten most important. 
\begin{center}
{\bf Ten Most Important Topics}
\end{center}
\begin{enumerate}
\item
What are the geometry and the topology of the universe? 
\item  
What are the bright standard candles that can be used in cosmology?
\item
What is the origin of primordial magnetic fields?
\item
What are the physical processes and phenomena associated with magnetic fields in the early universe?
\item
Besides the CMB, the LSS, SNIa, and GC, what other methods exist for measuring the cosmological parameters?
\item
What are dark matter and dark energy and what are their distributions in space and time?
\item 
Do primordial gravitational waves exist?
\item
Do ultrahigh energy cosmic rays (UHECR) have a cosmological origin?
\item
When, where and how did the first objects form? 
\item
Do we live in a universe of more than four dimensions?
\end{enumerate}
\par
In the following sections, we discuss each of the above topics. Throughout the text, we use 
$\hbar=c=k_{\rm B}=1.$
\section{WHAT ARE THE GEOMETRY AND THE TOPOLOGY OF THE UNIVERSE?}
\label{geom}
According to Einstein's equations, the universe can be hyperbolic
$(k=-1,\;\Omega_{\rm{T}}<1.0),$ flat $(k=0,\;\Omega_{\rm{T}}=1.000$...), or spherical
$(k=+1, \;\Omega_{\rm{T}}>1.0).$  If the value of $\Omega_{\rm{T}}$ is shown to be not exactly equal
to unity, the universe would then be hyperbolic or spherical. The observations of the CMB by the BOOMERANG 
balloon experiment indicate that $\Omega_{\rm{T}}=1.02 \pm 0.06$ when its data are taken in conjunction with 
that from SNIa and the LSS. The recent WMAP satellite data indicate that $\Omega_{\rm{T}}=1.02 \pm 0.02.$ It is possible that the universe started out with a value of $\Omega_{\rm{T}}$ appreciably different from unity. Due 
to an inflationary stage, it could have changed in value and become close to unity.
\par
However, it is important to look for other methods to determine whether $k=-1,0,$ or +1, instead of relying 
solely on the exact value of $\Omega_{\rm{T}}.$  Muller, Fagundes, and Opher \cite{MFO} showed that a universe created in a hyperbolic compact state could have a vacuum energy which is dependent on spatial position (Casimir Effect). This could, in principle, be observable. For example, the vacuum energy near us could be different from that near distant quasars.
\par
Although we generally assume that the universe is infinite, it could actually be finite(compact) and 
repeated many times. The finite size of the universe could be determined by the maximum spatial dimension of the density fluctuations created in the primordial universe, which could be measured by the CMB or the LSS. The 
recent WMAP CMB data have a smaller than expected quadropole and octopole, which may be an indication of a 
finite universe. One popular theory for the creation of the universe is based on quantum fluctuations. With such 
a mechanism, a small finite universe may have been easier to create.
\section{WHAT ARE THE BRIGHT STANDARD CANDLES THAT CAN BE USED IN COSMOLOGY?}
\label{sec:standard}
We need bright standard candles to study the universe at high redshifts. The brightest objects are SNIa,
quasars and Gamma Ray Bursts (GRB). Today, only SNIa are used in these studies. The problem with all
of these bright objects is that none of them is really a standard candle, i.e., none of them has a given
luminosity. In the case of SNIa, a correlation has been found between the peak luminosity and the width of the 
luminosity vs time curve for low redshifts. However, there is some doubt about whether they constitute good
standard candles at high redshifts.
\par
Quasars and GRB are brighter than SNIa and could, in principle, be used as standard candles at high
redshifts. To date, no correlation has been found between their luminosities and some intrinsic property, 
so as to enable them to be used as standard candles.
\subsection{\bf {SNIa}}
SNIa can attain $L_{\rm{GAL}},$ the luminosity of an entire galaxy. Although we base our present knowledge of 
the acceleration of the universe on the light curves of SNIa, we really don't know the source of the luminosity, i.e., whether it is due to the burning (deflagration) of a white dwarf with a Chandrasekhar mass, the burning of 
a white dwarf with less than this mass, or to the fusion of two white dwarfs.
\subsection{\bf {Quasars}}
Quasars can have luminosities $\sim 10^3\; L_{\rm{GAL}}$ and are observed at high redshifts. 
However, we do not understand the physical origin of their luminosities.
\subsection{\bf {GRB}}
GRB have peak luminosities which are $\sim 10^9\; L_{\rm{GAL}}$ (on the order of the luminosity of the entire universe) when they explode, making them observable at high redshifts. If we understood the physical processes involved in this phenomenon, we might be able to use them as standard candles at very high redshifts.
\section{WHAT IS THE ORIGIN OF PRIMORDIAL MAGNETIC FIELDS?}
\label{sec:origin}
We know that the microgauss magnetic fields in our galaxy are fundamental in the formation of stars. Since microgauss magnetic fields are observed in all galaxies at all redshifts, they may also be fundamental in the formation of the first objects. Evidence indicates that their origin cannot be due simply to the dynamo effect
since this requires a great many rotations and, at high redshifts, the galaxies could not have rotated a sufficient number of times. Furthermore, galaxies, such as the large and small Magellanic clouds, have been
observed to be both slowly rotating and to have microgauss fields, further indicating that the magnetic fields 
are primordial and do not owe their origin to the dynamo effect. Some possible origins of primordial magnetic fields are described below.
\subsection{\bf{Primordial Thermal Fluctuations}}
Small-scale primordial magnetic fields were created by thermal fluctuations, which can be  described by the Fluctuation-Dissipation Theorem (FDT). Black body radiation theory predicts zero amplitude, zero frequency 
fields. However, using FDT, which takes collective effects in the plasma into account, Opher and Opher \cite{OO1} predicted very large amplitude, zero frequency magnetic fields when the universe had a temperature 
$\sim\rm{MeV}.$  These fluctuations might not have been damped out completely during the evolution of the universe, but could have joined together (polimerized) to form large scale fields and continued to exist at lower temperatures until they formed the magnetic fields, observed in galaxies.
\subsection{\bf{Nonminimal Gravitational-Electromagnetic Coupling}}
Opher and Wichoski \cite{OW} suggested a theory of nonminimal gravitational-electromagnetic coupling, in which 
the magnetic dipole of an astrophysical object is proportional to its angular momentum (i.e., angular momentum produces magnetic fields). They found that the proportionality constant for the sun, moon, planets and pulsars is approximately the same to within an order of magnitude. Using the standard theory for the origin of angular momentum in galaxies, together with their theory of nonminimal gravitational-electromagnetic coupling, they found that the microgauss magnetic fields observed in galaxies are indeed produced. 
\subsection{\bf{Primordial Jets}}
Extragalactic jets are highly collimated and it has been suggested that this is due to an
axial current, producing a strong pinched magnetic field. They are also synchrotron radiation sources, which require magnetic fields, providing additional evidence for the existence of magnetic fields in jets. Since these jets have been seen at the highest observed redshifts, they probably also existed in the primordial universe. It is possible that jets in the early universe could have advected their magnetic fields into the primordial intergalactic medium, thereby giving rise to primordial fields, as has been suggested by Jafelice and Opher \cite{JO}. These fields could be the origin of the microgauss magnetic fields, observed in all galaxies today.
\subsection{\bf{Primordial Shocks}}
Some of the first objects to have collapsed are expected to have formed supernovae, which produce
strong shocks. The possibility of creating magnetic fields in primordial supernovae shocks was investigated by
Miranda, Opher and Opher \cite{MOO}. They assumed that the density gradients in the shocks were not exactly
parallel to the thermal gradients. It can be shown that when the density gradient in a plasma is not parallel to that of the temperature, a magnetic field is created. According to their theory, many successive supernovae 
shocks were created from a single supernova in the primordial universe, greatly amplifying the magnetic field, 
so that these fields could have been the origin of the microgauss  magnetic fields, observed in galaxies today. 
Accretion shocks are also expected to have been created in the formation of the first objects and it is possible that, in that case as well, the density gradients were not parallel to those of the temperature. 
\section{WHAT ARE THE PHYSICAL PROCESSES AND PHENOMENA ASSOCIATED WITH MAGNETIC FIELDS IN THE EARLY UNIVERSE?}
\label{sec:mag}
There are many important phenomena involving magnetic fields that took place in the early universe, about which, we still have little understanding, such as the formation of jets, accretion disks, and the production of relativistic electrons. 
\subsection{\bf{Formation of Jets}}
Jets have been discussed in the previous section in connection with their possible role in producing primordial intergalactic magnetic fields. The central role that jets play in the evolution of the universe can be seen from the following:
\begin{enumerate}
\item
We observe jets emerging from active galactic nuclei and quasars at all redshifts as well as from protostars.
\item
Jets are able to reach distances of over a Mpc (i.e., $\sim 100$ times greater than the size of a galaxy) and can transport energy, magnetic fields, and heavy elements as well as create turbulence. 
\end{enumerate}
\par
It is generally accepted that magnetic fields are needed in order to create jets (see Sec. IV). We also know 
that the magnetic fields in the jets must be regenerated. However, we do not yet understand either of these processes.
\subsection{\bf {Accretion Disks}}
The center of a galaxy (e.g., a central massive black hole) and a protostar are both examples of objects which become increasingly more massive due to the presence of an accretion disk. Accretion disks require a source of anamolous viscosity to accrete at the expected rates. This anamolous viscosity is generally attributed to 
magnetic fields, involving a process which is, as yet, not understood.
\subsection{\bf {Production of Relativistic Electrons}}
The presence of jets and magnetic fields in the primordial universe indicates the presence of synchrotron
radiation. Synchrotron radiation, observed in various astrophysical environments, including jets, is produced by 
relativistic electrons in magnetic fields. However, it is still not known exactly how the relativistic electrons are produced. 
\section{BESIDES THE CMB, the LSS, SNIa, AND GC,\\ WHAT OTHER METHODS EXIST FOR\\ MEASURING THE
BASIC COSMOLOGICAL PARAMETERS?}
\label{sec:BESIDES}
The basic cosmological parameters, such as the pressure of dark energy and the densities of dark matter and energy, are presently determined, primarily, by the CMB, the LSS, SNIa, and GC. Finding additional methods which could provide new information would be very important in order to cross-check results and put better limits on 
the uncertainties.  
\subsection{\bf Angular Distance vs Redshift}
The length of a standard candle at a given redshift defines an angle (called the angular distance) as a function
of the cosmological parameters. Although this method has been known for some time, an object that has a size 
that does not vary with redshift and, thus, can be used as a standard distance, has been 
difficult to find. One possibility is the length of compact jets from quasars $(\sim 20\; \rm{pc})$ since they 
are very small, compared to the size of the galaxy $(\sim 10\; \rm{kpc})$ \cite{LA}. Because of their small size, 
the physical processes which create the jets are probably intrinsic to the quasar and not to any possible interactions between the quasar and the intergalactic medium. Consequently, changes in the length of the jet 
could be correlated with other characteristics of the quasar, such as spectra, X-ray luminosity, etc. This is 
left for future research.
\subsection{\bf Distortion of a Sphere of Radius $R$ vs Redshift (Alcock-Paczynski Test)}
The line-of-sight radius $R_{\parallel}$ of a sphere as a function of redshift and the cosmological parameters
is seen differently than is the transverse radius $R_{\perp}$. This effect, measured by the Alcock-Paczynski 
Test, has been used to obtain the cosmological parameters by means of quasars, whose correlation function is 
known and which are bright enough to be observable at high redshifts \cite{CMW}. For an observed quasar at a 
given redshift, the correlation function is the probability of observing a second quasar at a distance $R$ 
from the first. The correlation function for quasars has a spherical form, $(R/R_0)^{-\gamma},$ where $R_0$ and $\gamma$ are known constants. Since the observed correlation function for $R_\perp$ is slightly different from that for $R_\parallel$ for a given $\gamma$ and $R_0,$ the Alcock-Paczynski Test predicts the observed cosmological parameters for a given $z$ as a function of the distortion of the correlation functions of $R_{\perp}$ and $R_{\parallel}.$ 
\subsection{\bf The Age of a Galaxy $\tau_{\rm{GAL}}$ vs Redshift}
The age of an elliptic galaxy provides a lower limit for the age of the universe at the redshift of the 
galaxy and can, therefore, be used to determine the cosmological parameters \cite{ALMA}. In a popular scenario 
for the formation of elliptic galaxies, most of the stars are assumed to have been formed at the same time.
For a star with a given mass, stellar evolution theory (SET) predicts the luminosity vs temperature as a 
function of time. Assuming a given initial mass function (IMF), (e.g., Saltpeter or Scalo), SET can predict the luminosity vs temperature of an entire galaxy with a given number of stars, as a function of time. Thus, from the observation of the luminosity vs temperature of a galaxy, we are able to determine its age, $\tau_{\rm{GAL}}.$
\par
A more accurate, but more laborious, method of obtaining $\tau_{\rm{GAL}}$ involves using the luminosity of a
galaxy as a function of frequency (using a spectrometer), instead of temperature (using an optical filter), which is an average over a range of frequencies.
\section{WHAT ARE DARK MATTER AND DARK ENERGY\\AND WHAT ARE THEIR
DISTRIBUTIONS IN SPACE AND TIME?}
\label{sec:dark}
The density of the universe is composed of $\sim 30\%$ dark matter and $\sim 70\%$ dark energy.
However, very little is known about the nature of either of these components.
\par
Known matter (baryonic matter) comprises only $\sim 4\%$ of the universe. The major part of dark matter, 
generally referred to as cold dark matter, interacts only gravitationally and has zero pressure. There is no 
known particle that behaves like cold dark matter. It has been suggested that cold dark matter is composed of neutralinos, the lightest supersymmetric particles. But until this particle with the correct properties
has been observed, it remains only a suggestion. 
\par
Dark energy behaves like vacuum energy, i.e., its pressure is approximately equal to minus its energy density.
If dark energy were indeed the vacuum energy, theory predicts that it should be $\sim 120$ orders of magnitude bigger than what is observed. A scalar field, known as quintessence, which has not been previously predicted
by particle physics theory, has been suggested as the source of the dark energy.
\par
It is to be noted how little we know about 96\% of the matter of the universe. We describe dark energy in terms 
of the vacuum energy, about which we know nothing, or by a scalar field, which has not been previously predicted by particle physics theory. The popular description of cold dark matter is in terms of a particle, which 
has not been observed.
\par
In quintessence, the potential of the scalar field needs to be fine-tuned, such that the quintessence energy density overtakes the matter density at the present epoch. In general, quintessence models have shallow 
potentials at the present epoch, requiring nearly massless fields. If coupling were to occur between the nearly massless field and ordinary matter, it could give rise to detectable long-range forces \cite{CAR}.
\par
Kamenshchik, Moschella and Pasquier \cite{KMP} assumed the existence of an exotic fluid, known as Chaplygin gas, to describe the transition from a universe which behaves as if it were composed of dust (when dark matter dominates) to an exponentially expanding universe (when dark energy dominates). The equation of state of the Chaplygin gas is $P=-A/\rho,$ where $P$ is the pressure, $\rho$ the energy density and $A$ is a positive 
constant. Chaplygin \cite{CHAP} introduced his equation of state in 1904 as a convenient solvable model for studying the aerodynamic lifting force on a plane wing. 
\par
Chaplygin gas may be an adequate description for plane lifting. However, the importance of its
role in cosmology has yet to be proven \cite{SAND}.
\section{DO PRIMORDIAL GRAVITATIONAL WAVES EXIST?}                                           
\label{sec:GRAV}
For a given potential that could have been capable  of creating the inflation era, $V(\phi)$, where $\phi$ is a 
scalar field called an inflaton, the creation of gravitational waves (tensor fluctuations) as well as density fluctuations (scalar fluctuations) are predicted by the Inflation Theory (see e.g., Lidsey et al. \cite{LID} for 
a review). Since the Inflation Theory was so successful in predicting  the approximate scale-invariant 
distribution of density fluctuations, which is, indeed, observed in the CMB and the LSS, it is to  be hoped that it will have equal success in predicting the gravitational wave spectrum, which requires data of a higher precision, not yet available. Precision spectra of the CMB are expected to be obtained in the near future, so 
that a better comparison with predictions can be made. These precision spectra of the CMB may also provide information about the possible creation of gravitational waves in phase transitions (e.g., quark-hadron, electroweak, etc.) in the primordial universe.
\par
In the formation of the first objects (discussed in Section X), a certain fraction of the mass of the first objects formed black holes, emitting gravitational waves in the process. These waves could be detected by a sufficiently sensitive pair of gravitational wave detectors, such as LIGOs (Laser Interferometer Gravitational Observatories), which are under construction, or LISA (Laser Interferometer Space Antenna), which is in the planning stage. 
\section{DO UHECR HAVE A COSMOLOGICAL ORIGIN?}
\label{sec:UHECR}
At the present time, the origin of UHECR $>10^{20}\;\rm{eV}$ remains a mystery. Whereas their isotropic distribution indicates a cosmological origin, the GZK cutoff points to a local origin $(<50\;\rm{Mpc\; h^{-1}},$ where $h=H_0/100\;\rm{km\;s^{-1}\;M_{pc}^{-1}}$). It is generally assumed that UHECR are protons. The GZK cutoff 
is due to the strong damping of protons $>10^{20}\;\rm{eV},$ which have damping lengths of $\sim 50\;\rm{Mpc\; h^{-1}},$ as a result of their interaction with CMB photons. However, the distribution of strong radio galaxies, which could be the astrophysical sources of UHECR, is not isotropic within $50\;\rm{Mpc\;h^{-1}}.$ In fact, 
within this distance, there is only one powerful radio source (in the Virgo cluster) that could be the origin of UHECR. One way to avoid the GZK cutoff and, at the same time, explain the approximately isotropic distribution, 
is to assume that UHECR are due to the decay of extremely massive particles $(m>10^{20}\;\rm{eV})$ 
in our halo, which were produced in the primordial universe. These particles could, in principle, have been created in the period of the post-inflation era, when reheating took place, along with 
all the other known matter. However, in that period, the temperature is thought to have been 
$\ll 10^{20}\;\rm{eV},$ making it impossible for these extremely massive particles to have been created then. Therefore, their creation must have occured in the post-inflation era, but before the beginning of the reheating era, for which various mechanisms have been suggested (see for example Chung, Kolb, and Riotto \cite{CKR}).
\section{\bf WHEN, WHERE, AND HOW DID THE FIRST OBJECTS FORM?}
\label{sec:FIRST}
The recent WMAP CMB data indicate that the first objects formed at a redshift $\sim 20.$ To understand the formation of the first objects, we must understand the physics of collapse, production of relativistic 
particles, effects of turbulence, and the formation of low-mass objects, among other important processes at a redshift $\sim 20$, when the universe had negligible metalicity (i.e., no dust for cooling) and 
the average baryon density of the universe was $\sim 9,300$ times greater than it is today.
\subsection{\bf Collapse}
In the past, studies of the collapse of a cloud, out of which, the first objects were formed, were primarily 
based on gravitational effects. There are, however, important plasma effects which must be taken into account
when treating collapse after the recombination era. The introduction of plasma effects in the treatment
of collapse after the recombination era was first begun by de Araujo and Opher in 1988 \cite{AO} and 
continued later by Haiman, Rees, and Loeb in 1996 \cite{HRL} and Tegmark et al. in 1997 \cite{TEG}.
\par
Plasma effects could have made themselves felt by aiding or interfering with the collapse process of a cloud 
in a variety of ways.
\begin{enumerate}
\item{Photon-drag}, the scattering of electrons by CMB photons in the collapsing cloud, would have slowed 
down the collapse process. 
\item{Photon cooling}, the scattering of the electrons by the cooler CMB photons in the cloud, would have 
acted to prevent the heating of the cloud, thus aiding the collapse process.
\item{The formation (destruction) of $\rm{H_2}$ molecules},
the prime primordial coolant at $T<10^4\;\rm{K}$ and essential for the collapse of objects, would have aided (interfered with) the collapse process.
\item{Primordial magnetic field effects} (see Section IV) would have been felt in two ways: (1) magnetic pressure acting to prevent collapse; (2) emission of torsional Alfv\'en waves or rotating jets, reducing the angular momentum of the first objects and, thereby, aiding the collapse process.  
\item{A thermal instability} creating
a two-phase region in pressure equilibrium, with one region at a high temperature and a low density and the other region at a low temperature and a high density, could have aided the collapse process. If the low 
temperature-high density region were to have had a mass greater than the Jeans mass, it would have 
gravitationally collapsed.
\end{enumerate}
\subsection{\bf{Production of Relativistic Particles}}
Particles were accelerated in the early universe by shocks due to: (1) primordial explosions (e.g., primordial supernovae); (2) accretion in the formation of the first objects; or (3) the collision of primordial galaxies. Galactic supernovae shocks are the accepted source of cosmic rays of energies $\leq 10^{15}\;\rm{eV}.$ In the linear treatment of shocks, in which the pressure of the accelerated particles is neglected, a power-law energy distribution of cosmic rays is predicted, as is, in fact, observed. However, the linear treatment predicts the growth of the particle pressure until it becomes greater than that of the gas for strong shocks, which is clearly
impossible. Thus, a non-linear theory, which includes the pressure of the accelerated particles as well as that 
of the Alfven waves, which scatter the particles, is required. In a non-linear treatment of strong shocks, Medina-Tanco and Opher \cite{MT} showed that more than 50\% of the kinetic energy of the gas can be transfered to the accelerated particles. The radiation from the accelerated particles produced by the primordial shocks could, in principle, be observed in the microwave, infrared, X-ray, and gamma-ray background radiation.
\subsection{\bf Turbulence}
The effect of turbulence in the primordial universe is particularly important since it can transfer energy 
from small to large scales in, what is known as, an inverse energy cascade, which may have been important in the
formation of the first objects. Studying the turbulence in a nearby large HII extragalactic cloud
(radius $\sim 80\;\rm{pc}$), within which, there are a few very massive hot stars heating up small volumes,  
Medina-Tanco et al. \cite{MSJO} showed, for the first time in an object outside of the solar system, evidence
for an inverse energy cascade. In this case, energy deposited in a region $\sim 10\;\rm{pc}$ by the massive hot stars is being transferred to the edge of the cloud, a region at a distance $\sim 80\;\rm{pc}$ away. In the process of the collapse of large clouds to form the first objects, large amounts of energy also heated up small regions. If inverse energy cascades were to have formed, energy would have been transferred outwards, so that turbulence would have played an important role in energy transfer during the formation of the first objects.
\subsection{\bf Formation of Low-Mass Objects}
Part of the first objects formed could have had low masses ($< M_{\odot}$). Such objects, whose formation is yet to be understood, include MACHOS (MAssive Compact Halo Objects) and planetary objects.
\par
Using micro-gravitational lensing, MACHOS, which do not emit light, appear
to have masses $\sim 0.5\; M_\odot$ \cite{ALCO}. An object $\sim 0.5\; M_\odot$ that does not
emit light could, in principle, be a dead white dwarf. However, the number of MACHOs is $\sim 100$ times that of 
expected dead white dwarfs. Furthermore, such a large number of dead white dwarfs would have produced many more heavy elements during their lifetimes than are, in fact, observed. The only known $0.5 M_{\odot}$ object that passed through stellar evolution and no longer emits light is a dead white dwarf. Thus, MACHOs appear to have 
been formed without passing through the stage of stellar nuclear burning. One possibility is that MACHOS's are black holes, formed in the primordial universe. Another possibility is that they are very massive planets, 
formed by a thermal instability in the primordial universe.
\par
The formation of isolated massive planets is just as mysterious as is the formation of MACHOs.
Zapatero et al. \cite{ZAP} discovered isolated planetary objects $\sim 5$ Jupiter masses, which are not 
connected to any stars. 
\section{\bf DO WE LIVE IN A UNIVERSE WITH MORE THAN FOUR DIMENSIONS?}
\label{sec:FOUR}
Various theories of unification involve more than four dimensions. Accordingly, cosmology theories with more 
than four dimensions have also appeared . 
\par
The Planck energy, $M_{\rm{PL}}\sim 10^{19}\;\rm{GeV}$, at which gravitational quantum effects become important
and the gravitational interaction becomes comparable to the electroweak interaction \cite{ARK}, is very much greater than $\sim \rm{TeV}$, the unification energy of the electroweak interaction. Introducing extra 
dimensions allows for the possibility of reducing the value of the Planck energy. Therefore, a theory of a universe with more than 4 dimensions was created, in which the new Planck energy is 
$M_{\ast\rm{PL}}\sim\rm{TeV},$ thus achieving the first step in the unification of gravitation with the electroweak interaction.       
\par
In this theory, our 4-D world, called a brane (from the word ``membrane"), would be a part of such a universe. 
The existence of the rest of the universe would not be perceived as long as we are observing non-gravitational interactions, which take place in four dimensions. However, gravitation is the one interaction that can 
penetrate into the extra dimensional part of the universe and interact with it. This interaction enables us to 
be made aware of the rest of the universe outside of our brane.
\par
The value of the new Planck energy is directly related to the number and size of the extra dimensions. If $R_n$ 
is the size of the $n$ extra dimensions, the relation between $M_{\rm{PL}},$ $R_n,$ and $M_{\ast\rm{PL}}$ in 
this theory is $(M_{\rm{PL}}/M_{\ast\rm{PL}})^2\sim (R_n M_{\ast\rm{PL}})^{\rm{n}}.$ For distances $\gg R_n,$ 
we use Newton's law of gravitation, but modify it for distances $\ll R_n.$  Thus, in order to achieve 
unification at the TeV energy with one extra dimension, we set $M_{\ast\rm{PL}}\sim 1\;\rm{TeV}$ and, 
using $n=1,$ obtain $R_1=10^{13}\;\rm{cm}$ from the above relation. However, this value of $R_n$ with $n=1$ violates observations in the solar system, which are correctly described by Newton's law. Since $M_{\rm{PL}}$ cannot be reduced to 1 TeV by means of only one extra dimension, an attempt at using two extra dimensions is currently being investigated in the laboratory.  With $M_{\rm{PL}}\sim 1\;\rm{TeV}$ and $n=2,$ we obtain 
$R_2\sim 1\rm{mm}.$ Thus, we find that this extra dimension theory with $n=2$ predicts a modification of the gravitational interaction for very small distances. 
\par
Such a modification can affect nuclear reactions in cosmology and astrophysics as well as in the laboratory, 
when nucleon-nucleon bremsstrahlung with graviton (instead of photon) emission is involved. Gravitons are 
produced by nucleon-nucleon bremsstrahlung, which has a temperature dependent cross-section $\propto (T/M_{\ast\rm{PL}})^{n+2},$ where $T$ is the temperature of the medium. Upper limits on graviton production 
from observations put restraints on the theory.
\par
Applying this theory to Big Bang cosmology, in which the early universe was extremely hot, 
we must require that above a certain temperature $T_\ast,$ the extra dimensions be empty of energy density 
in order to avoid the production of too many gravitons. For $M_{\ast\rm {PL}}\sim 1\; \rm{TeV}$ and $n=2$ 
(worst case), we require that $T_\ast\leq 10\;\rm{MeV},$ so that the expansion rate due to the extra gravitons does not violate observations of the primordial nucleosynthesis of the light elements D, $^3$He, $^4$He, and $^7$Li, which occurred at $T\sim 1-0.01\;\rm{MeV}.$ 
\par
The decay of gravitons into photons can affect the background photon radiation. This puts a limit on the production of primordial gravitational radiation, predicted in this theory of extra dimensions. Gravitons
have an effective mass $T_\ast$ and a lifetime {$\propto T_\ast^{-3}$} to decay into photons.
For example, in order not to violate the 1 MeV ($\sim T_{\ast}$) photon background, we require that for $n=2$, $M_{\ast\rm{PL}}\geq 10\;\rm{TeV}.$ 
\par
Production of gravitons in extra dimensions could violate what we know about the sun. Nuclear reactions at the
center of the sun produce a luminosity which is in approximate agreement with stellar evolution theory.
Therefore, a large fraction of the energy produced by nuclear reactions cannot go into graviton production
since this does not contribute to the optical luminosity. The sun releases energy at a rate 
$\varepsilon_s\sim 1\;\rm{erg\; gm^{-1}\;s^{-1}}.$ In the nucleon-nucleon bremsstrahlung cross-section, 
$\propto (T/M_{\ast\rm{PL}})^{n+2},$ where $T$ is the central temperature of the sun, we require that
$M_{\ast\rm{PL}}> 30\;\rm{GeV}$ for $n=2,$ so that the cooling rate due to graviton emission is 
$\ll\varepsilon_s.$ 
\par
Graviton production could also affect what we know about supernovae, where the central temperature is higher, making the limits imposed on this theory even stronger. For a supernova, such as SN1987a, the central 
temperature reaches $T\sim 30\;\rm{MeV}.$ In order for graviton production not to affect the optical
luminosity, we require that for n=2, $M_{\ast\rm{PL}}> 30\;\rm{TeV}.$ 
\par
Present laboratory accelerators produce center-of-mass energies $E_{\rm{CM}}>1\;\rm{TeV},$ putting strong
limits on $M_{\ast\rm{PL}}$ \cite{LAN}. Effects of $E_{\rm{CM}}>M_{\ast\rm{PL}}$ could appear in two ways in
accelerator experiments: 
\begin{enumerate}
\item                                                                                         
Due to graviton production, an apparent non-conservation of energy could occur.
\item
Due to the effect of virtual gravitons in the interactions, an anomalous production of
fermion-fermion pairs and/or diboson pairs could occur.
\end{enumerate}
Abbott et al. \cite{ABB} found that $M_{\ast\rm{PL}}>1.0-1.4\;\rm{TeV}$ for $E_{\rm{CM}}\sim 1.8\;\rm{TeV}.$  
Therefore we will have to wait for accelerators with higher $E_{\rm{CM}},$ in order to try to determine the 
value of $M_{\ast\rm{PL}}.$ For $E_{\rm{CM}}\gg M_{\ast\rm{PL}},$ black holes and branes should be created in 
the laboratory and in the interaction of UHECR with the atmosphere \cite{CAV}. An example of brane production 
was given by Jain et al. \cite{JKP}. They suggested that neutrinos of energy $10^{11}\;\rm{GeV}$ have very large cross sections due to the production of branes and that they could explain the observed UHECR (see Section IX).
\par                                                        
Whereas the previous extra dimension theory dealt with modifications of short-range interactions (nuclear reactions at small distances (high energies)), we can try to create a different extra dimension theory (e.g., a 5-D theory with $n = 1$), in which long-range interactions (gravitational) are modified, so as to treat the dark energy problem. In this new theory, the extra dimensions are felt in the region where $R > R_1 \sim H_0^{-1}$\cite{DDG}.  Here, the extra dimensions extend to infinity and are felt for distances $> R_1$. At such distances, the gravitational potential does not have a $1/R$ dependence (as in newtonian theory), but decreases 
as $1/R^2.$
\par
Distances were $\ll R_1\sim H^{-1}_0$ in the distant past, when the universe was much smaller. At that time, 
the normal newtonian gravitational interaction was valid, according to this theory. Now that the universe has 
a size $\sim R_1,$ the gravitational potential energy is becoming weaker due to the penetration of gravitation into the extra dimension and, as a result, the matter in the universe is gaining kinetic energy and accelerating. In Section VII, we discussed the introduction of dark energy into cosmology to explain the observed acceleration of the universe. Here, the acceleration is explained by the existance of extra dimensions. 
\par
In this new extra dimension theory, the cosmological constant (vacuum energy) is taken to be zero. The new 
Einstein equation for the Hubble parameter as a function of $z$ has the form
$H^2/H^2_0=\Omega_k (1+z)^2+{\{\Omega^{1/2}_{\rm{1}}+{[\Omega_{\rm{1}}+\Omega_M(1+z)}^3]^{1/2}\}}^2,$
instead of $H^2/H_0^2=\Omega_k (1+z)^2 + \Omega_M(1+z)^3$ in the standard theory form, where 
$\Omega_{\rm{1}}\equiv 4{R_1}^2 {H_0}^2$ and $\Omega_k=-k/{H_0}^2 \;\;(k=+1,\; 0,\; -1)$.  For large redshifts (in the distant past), the two equations are identical.
\par
This 5-D theory can be tested by studying light distances to standard candles (e.g., SNIa). Comparing light
distances in this theory with those in the popular quintessence models, in which $\Omega_x=0.7,\;\Omega_M=0.3,$ and $w_x\equiv P_x/ \rho_x>-1$ (see Section VII), it is found that for $z\sim 2-4,$ the light distance is $\sim 9\%$ greater in the 5-D theory than in the quintessence model for $w_x=-0.6$ and $\sim 3\%$ greater than in the quintessence model for $w_x=-0.8.$ 
\par
A comparison between the 5-D theory and the SNIa data, was also made by Deffayet et al. \cite{DDG}. They found that $R_1=1.2{H_0}^{-1}$ for $\Omega_M=0.3$ and that the beginning of the acceleration of the universe 
took place at $z_A=0.8.$ In agreement with Alcaniz \cite{ALZ}, they found that acceleration started later 
(smaller $z$) in the 5-D model than in quintessence or vacuum energy models (Section VII).
\par
Deffayet et al. \cite {DLR} found that the 5-D theory can be made to resemble a quintessence model (i.e., a
4D model) by the introduction of an effective $w_x(z).$ This $w_x(z)$ varies with redshift as $w_x\simeq - (1+\Omega_M)^{-1}\;\simeq - 3/4$ for small $z$ and as $w_x\simeq - 1/2$ for large $z$. 
\par
Based on a theory of extra dimensions, a cyclical cosmological model of the universe, with the
universe oscillating forever between big crunches-big bangs and maximum expansions, was suggested by
Steinhardt and Turok \cite{ST}. This extra dimension model can be described in 4 dimensions, with an effective scalar field $\phi$ representing the penetration of gravitation into other dimensions \cite {ST}. The $\phi$ 
field has a potential, $V(\phi)\propto(1-{\rm{e}}^{-C\phi}),$ where $C$ is a constant and the form ${\rm{e}}^{-C\phi}$ comes from extra dimension theories. This model also includes a function $\beta(\phi)$ that
describes the interaction of $\phi$ with matter and radiation. 
\par
Thus, it appears that many different greater than 4-D theories can be created with the justification that 
they are implied by extra-dimension theories of unification. However, it should be emphasized that, at the 
present time, no extra-dimension unification theory actually exists. It also should be noted that, although 
extra dimensions may be necessary in a complete theory of unification, they may not be needed in cosmology 
(e.g., it is not necessary to use general relativity to describe the trajectory of a rocket; newtonian theory 
is sufficient).
\section{SUMMARY}
\label{sec:SUM}
This lecture focused on some of what I consider to be the most important problems in cosmology today.
We are still very far from the precision era of cosmology. No less than 96\% of the universe is 
completely unknown to us. Existing cosmological theories are not able to fully explain the existing 
observational data and must be viewed as interesting and imaginative scenarios. The observations, 
themselves, are insufficient to provide a clear picture of just how the universe
evolved after the Big Bang. We will have to wait for better observations in the future
to help us develop a better theory, against which, we can test predictions.
\section{ACKNOWLEDGEMENTS}
\label{sec:ACK}
The author would like to thank 
the Brazilian agencies FAPESP (Proc. No. 00/06770-2), FINEP 
(Pronex No. 41.96.0908.00), and CNPq for partial support.

\end{document}